# SUSY Dark Matter in Nonuniversal Gaugino Mass Models

## D. P. Roy

Homi Bhabha Centre for Science Education, Tata Institute of Fundamental Research, Mumbai 400088, India

e-mail: dproy@tifr.res.in

**Abstract:** We discuss the SUSY dark matter phenomenology in some simple and predictive models of nonuniversal gaugino masses at the GUT scale. Assuming the gaugino masses to transform as a sum of singlet and a nonsinglet representation of the GUT group SU(5), one can evade the LEP constraints to access the bulk annihilation region of the bino dark matter relic density. Besides, with this assumption one can also have a mixed gaugino-higgsino dark matter, giving the right relic density over large parts of the parameter space. We consider the model predictions for LHC and dark matter experiments in both the cases. Finally we consider the AMSB model prediction of wino dark matter giving the right relic density for TeV scale wino mass. Assuming this wino dark matter mass to be at the first Sommerfeld resonance of ≈ 4 TeV one can simultaneously reproduce the right relic density as well as the hard positron spectrum observed by the PAMELA experiment.

## 1. Introduction

The most phenomenologically attractive feature of SUSY and in particular the minimum supersymmetric standard model (MSSM) is that it provides a natural candidate for the dark matter (DM) in terms of the lightest neutralino $\tilde{\chi}_1^0$ (abbreviated as $\chi$). In the constrained version of this model (CMSSM), corresponding to universal gaugino and scalar masses at the GUT scale, the dark matter is dominantly bino. And since the bino carries no gauge charge, its main annihilation mechanism is via sfermion exchange. This is called the bulk annihilation process; and the region of parameter space giving the right DM relic density via this process is called the bulk annihilation region. Unfortunately, the LEP limit on Higgs boson mass imposes rather stringent limits on bino and sfermion masses in the CMSSM, which rules out the bulk annihilation region of its parameter space [1]. Therefore the model can give the right DM relic density only over a few narrow strips of parameter space, corresponding to fairly high degrees of fine-tuning.

We consider here the DM phenomenology of a set of simple and predictive models of nonuniversal gaugino masses, which are assumed to transform like the sum of the singlet and a nonsinglet representation of the GUT group SU(5). With this assumption, one can evade the LEP constraint to access the bulk annihilation region of a bino DM. Besides one can also have a mixed bino-higgsino DM giving the right relic density over a large part of the parameter space. In section 2 we shall introduce the basic model assumptions and these two types of DM models resulting from it. Then in sections 3 and 4 we shall discuss the implications of these two types of DM models for LHC and DM experiments. Finally in section 5 we shall discuss the AMSB model prediction of a TeV scale wino DM along with its implications for DM experiments.

## 2. Nonuniversality of Gaugino Masses in SU(5) GUT

The gauge kinetic function responsible for GUT scale gaugino masses arises from the vacuum expectation value of the F term of the chiral superfield $\Phi$ responsible for SUSY breaking,

$$\frac{\langle F_\Phi \rangle_{ij}}{M_{Plank}} \lambda_i \lambda_j, \tag{1}$$

where $\lambda_{1,2,3}$ are the U(1), SU(2) and SU(3) gaugino fields – bino, wino and gluino. Since the gauginos belong to the adjoint representation of the GUT group SU(5), the $\Phi$ and $F_\Phi$ can belong to any of the irreducible representations appearing in their symmetric product,

$$(24 \times 24)_{symm} = 1 + 24 + 75 + 200. \tag{2}$$

Thus the GUT scale gaugino masses for a given representation of the SUSY breaking superfield $\Phi$ are given in terms of a single mass parameter [2],

$$M^G_{1,2,3} = C^n_{1,2,3} m^n_{1/2}, \tag{3}$$

where

$$C^1_{1,2,3} = (1,1,1); C^{24}_{1,2,3} = (-1,-3,2); C^{75}_{1,2,3} = (-5,3,1); C^{200}_{1,2,3} = (10,2,1). \tag{4}$$

The CMSSM assumes $\Phi$ to be a singlet, implying universal gaugino masses at the GUT scale. On the other hand, any of the nonsinglet representations of $\Phi$ would imply nonuniversal gaugino masses via (3) and (4). These nonuniversal gaugino mass models are known to be consistent with the universality of gauge couplings at the GUT scale, with $\alpha_G \approx 1/25$; and their phenomenology has been widely studied [3]. The superparticle masses at the EW scale are related to these GUT scale gaugino masses and the universal scalar mass parameter $m_0$ via the RGE. In particular the gaugino masses evolve like the corresponding gauge couplings at the one-loop level of the RGE, i.e.

$$M_1 = (\alpha_1/\alpha_G) \simeq (25/60) C_1^n m_{1/2}^n,$$
$$M_2 = (\alpha_2/\alpha_G) \simeq (25/30) C_2^n m_{1/2}^n, \qquad (5)$$
$$M_3 = (\alpha_3/\alpha_G) \simeq (25/9) C_3^n m_{1/2}^n.$$

The corresponding higgsino mass μ is obtained from the EW symmetry breaking condition along with the one-loop RGE for the Higgs scalar mass. i.e.

$$\mu^2 + M_Z^2/2 \simeq -m_{H_u}^2 \simeq -0.1 m_0^2 + 2.1 M_3^{G^2} - 0.22 M_2^{G^2} - 0.19 M_3^G M_2^G \qquad (6)$$

neglecting the contribution from the GUT scale trilinear coupling term $A_0$ [4]. The numerical coefficients on the right correspond to a representative value of tan β = 10; but they show only mild variations over the moderate tan β region.

Although our results are based on exact numerical solutions to the two-loop RGE [5], the approximate formulae (5) and (6) are very useful in understanding the essential features of the models. Combining these formulae with (4) shows that for the 1 and 24-plet representations μ > $M_1$ implying bino DM, while for the 75 and 200-plet representations μ < $M_1$ implying a higgsino DM. As mentioned earlier, the bino DM generically gives an over-abundance of relic density except for the bulk annihilation region. On the other hand, the higgsino DM can efficiently annihilate via their isospin gauge coupling, leading to an under-abundance of relic density for sub-TeV DM mass. (A higgsino DM mass of about 1 TeV gives the right relic density, but will be out of reach of LHC.) One gets phenomenologically promising DM models by assuming a combination of two SUSY breaking superfields belonging to singlet and one of the above nonsinglet representations of SU(5). It was shown in [6] that one can evade the above mentioned LEP constraint to access the bulk annihilation region of right DM relic density for the 1+24, 1+75 and 1+200 representations. Moreover it was shown in [7] that one can also get mixed bino-higgsino DM for the 1+75 and 1+200 representations, leading to right relic density over large parts of the parameter space. We shall discuss the implications of these two types of models for LHC and DM detection experiments in the next two sections.

## 3. Bino DM in the Bulk Annihilation Region in 1+75 and 1+200 Models

For the bulk annihilation region in the 1+75 and 1+200 models the singlet contributions to gaugino masses dominate over the nonsinglet ones, while it is the other way around in the 1+24 model [6]. We shall concentrate here on the first two models. For a general nonuniversal gaugino mass model the bulk annihilation region of bino DM spans over the parameter range

$$M_1^G = 150 - 250 \, GeV \ \& \ m_0 = 50 - 80 \, GeV, \tag{7}$$

practically independent of the other SUSY parameters [8]. This is because the main annihilation process for the bino DM pair proceeds via right handed slepton exchange,

$$\chi\chi \xrightarrow{\tilde{l}_R} \bar{l}l\,; \tag{8}$$

and the bino mass is determined by $M_1^G$ via the RGE (5), while the right handed slepton mass is determined via its RGE by $M_1^G$ and $m_0$. For our models we have one more gaugino mass parameter at our disposal. We choose this to be $M_3^G$, because it makes the dominant contribution to the squark/gluino masses via their RGE as well as the higgsino mass via (6).

$M_1^G = 250$ GeV, $M_3^G = 800$ GeV, $m_0 = 67$ GeV

| Particle | Mass (GeV) | |
|---|---|---|
| | (1+75) model | (1+200) model |
| $\tilde{\chi}_1^0$ (bino) | 100 | 99.6 |
| $\tilde{\chi}_2^0$ (wino) | 772 | 586 |
| $\tilde{\chi}_3^0$ (higgsino) | 933 | 970 |
| $\tilde{\chi}_4^0$ (higgsino) | 955 | 979 |
| $\tilde{\chi}_1^+$ (wino) | 772 | 586 |
| $\tilde{\chi}_2^+$ (higgsino) | 955 | 979 |
| $M_1$ | 102 | 102 |
| $M_2$ | 778 | 579 |
| $M_3$ | 1718 | 1723 |
| $\mu$ | 928 | 965 |
| $\tilde{g}$ | 1766 | 1766 |
| $\tilde{\tau}_1$ | 100 | 104 |
| $\tilde{\tau}_2$ | 627 | 474 |
| $\tilde{e}_R, \tilde{\mu}_R$ | 119 | 119 |
| $\tilde{e}_L, \tilde{\mu}_L$ | 628 | 474 |
| $\tilde{t}_1$ | 1221 | 1251 |
| $\tilde{t}_2$ | 1541 | 1507 |
| $\tilde{b}_1$ | 1512 | 1480 |
| $\tilde{b}_2$ | 1529 | 1528 |
| $\tilde{q}_{1,2,R}$ | ~1528 | ~1533 |
| $\tilde{q}_{1,2,L}$ | ~1640 | ~1593 |

Table1. SUSY mass spectrum in the 1+75 and 1+200 models for the bino DM mass near the upper end of the bulk annihilation region and a representative value of tan β = 10 [9].

Table 1 shows the SUSY mass spectrum in the 1+75 and 1+200 models for the bino DM mass near the upper end of the bulk annihilation region, where we have chosen $M_3^G$ = 800 GeV [9]. The resulting values of squark/gluino masses in the range of 1500-1700 GeV are outside the reach of the 7-8 TeV LHC, but well within that of 14 TeV LHC data. Note that the $M_2^G$ values for the two models are determined via (3) and (4). Therefore the resulting values of wino and left slepton masses are different for the two models. However, the small value of $m_0$ ensures that the left sleptons are lighter than the wino for each model. Consequently the models predict the SUSY cascade decay at LHC to proceed via these slepton states resulting in a distinctive signal with a hard electon/muon or tau-jet accompanying the missing-$E_T$.

The direct DM detection experiments are based on their elastic scattering on a heavy nucleus like Germanium or Xenon, which is dominated by the spin-independent $\chi p$ scattering mediated by the Higgs boson exchange. Since the Higgs coupling to the lightest neutralino $\chi$ is proportional to the product of its higgsino and gaugino components, the direct detection cross-section is expected to be small for a bino dominated $\chi$ state. Numerical result using DarkSUSY [10] gives a cross-section well below the current experimental limits [11-13], but within the reach of a future 1 ton Xenon experiment [14].

The indirect DM detection experiments are based on observing the decay products of their pair annihilation at the present time. Since the DM particles are highly non-relativistic (v ~ $10^{-3}$), only the S-wave annihilation cross-sections are of any significance at the present time. One can then show from symmetry considerations that for Majorana particles like the neutralino $\chi$ the cross-section for the annihilation process (8) is helicity suppressed by a factor of $(m_l / M_W)^2$. In contrast the cross-section for the radiative annihilation process

$$\chi\chi \xrightarrow{\tilde{l}_R} \bar{l}l\gamma \tag{9}$$

is only suppressed by a factor $\alpha$, so that it constitutes the dominant annihilation process at the present time [15]. We have computed the resulting hard positron signal for the SUSY spectrum of table 1 using the DarkSUSY [10] and Galprop [16] codes. As expected the result depends only on the input parameters $M_1^G$ and $m_0$ irrespective of the other model parameters. Hence it provides a robust prediction for the bino DM models in the bulk annihilation region [9]. The figure 1 compares the model prediction with the hard positron spectrum observed by the PAMELA experiment [17]. There is good agreement between the shape of the predicted spectrum with this data, with only the last data point overshooting the prediction by two standard deviations. Note that the annihilation process (9) produces no antiprotons and hence predicts no antiproton excess over the cosmic ray background in agreement with the PAMELA data. However, the model has no explanation for the huge boost factor of ~ 7000 required to match the size of the PAMELA positron signal. Therefore one needs to attribute this factor to astrophysical sources like a local population of intermediate mass black holes leading to spikes in the DM density distribution [18], or a nearby DM clump [19].

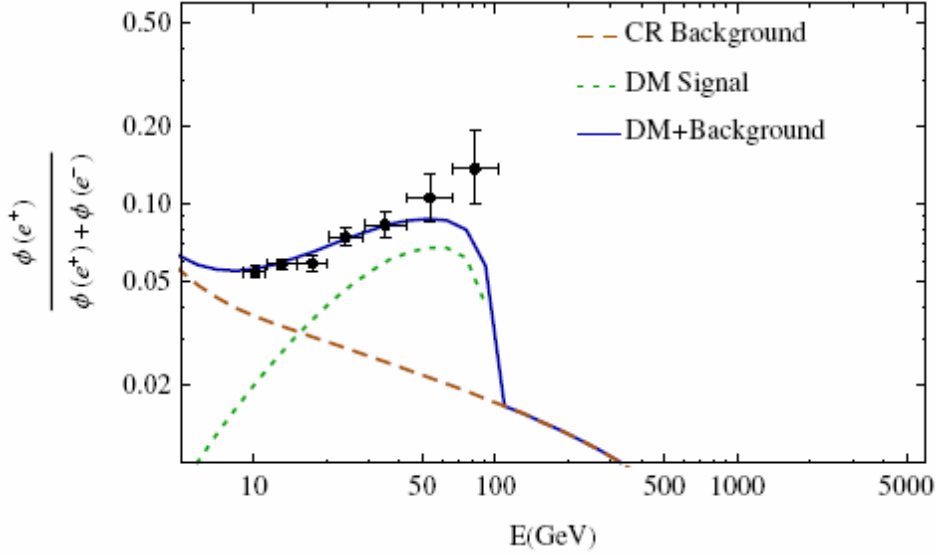

Fig 1. The predicted positron flux of the bino DM in the bulk annihilation region is compared with the PAMELA data assuming a boost factor of ~ 7000 [9].

## 3. Mixed Neutralino DM in the 1+75 and 1+200 Models

In these models the relative values of the singlet and nonsinglet gaugino mass parameters of (3) can be described in terms of a dimensionless parameter α [7], i.e.

$$m_{1/2}^1 = (1-\alpha_{75})m_{1/2}, m_{1/2}^{75} = \alpha_{75}m_{1/2} \ \& \ m_{1/2}^1 = (1-\alpha_{200})m_{1/2}, m_{1/2}^{200} = \alpha_{200}m_{1/2}. \tag{10}$$

Then one sees from (3-6) that for the 1+75 model,

$$\alpha_{75} = 0.5 \Rightarrow |M_1| \simeq |\mu| \simeq m_{1/2} < |M_2| \text{ for } m_0 \geq m_{1/2}. \tag{11}$$

This corresponds to a mixed neutralino DM with bino and higgsino components of roughly similar size. It represents a simple realization of the so called well-tempered neutralino scenario [20]. It is analogous to the mixed bino-higgsino DM in the focus point region of the CMSSM [21], except that it holds over a wide region of the parameter space here. Figure 2 shows the $m_0$-$m_{1/2}$ parameter space of the 1+75 model for an optimal choice of $\alpha_{75}$ = 0.475 with contours of $M_1$ and μ. It also shows the contours of the bino component of DM and the DM relic density. The regions satisfying the cold DM relic density measured by WMAP [22]

$$\Omega h^2 = 0.110 \pm 0.019(3\sigma) \tag{12}$$

are indicated by the bands of red dots. The upper band corresponds to the DM being an almost equally mixed bino-higgsino state, where the gauge coupling of the higgsino component leads to the pair annihilation of χχ into WW, ZZ and fermion-antifermion pairs (via Z) as in the focus point region of the CMSSM. On the other hand, the lower band represents the upper edge of the resonant annihilation region of $\chi\chi$ into a heavy quark(lepton) pairs via the pseudo scalar Higgs boson A. The lower edge appears as a narrow strip at the lower right corner, while the region in between has under abundance of DM relic density due to efficient resonant annihilation. Thus the lower branch enclosed by these two edges is similar to the so called funnel region of the the CMSSM. But since the higgsino component of ~ 10% here is still significantly larger than in CMSSM, the funnel region occurs even for a moderate $\tan\beta = 10$ and spans a wide region of parameter space.

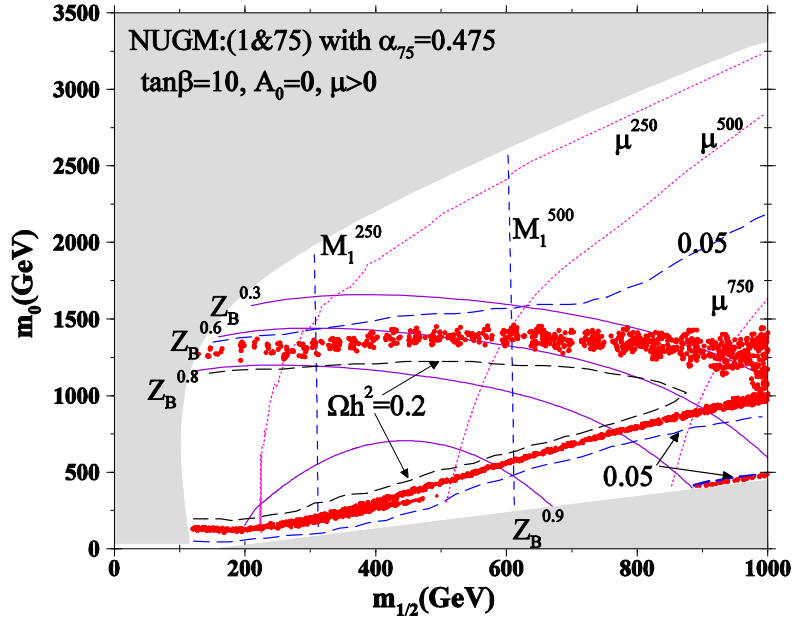

Fig 2. The composition and relic density of the mixed netralino DM of the 1+75 model [7].

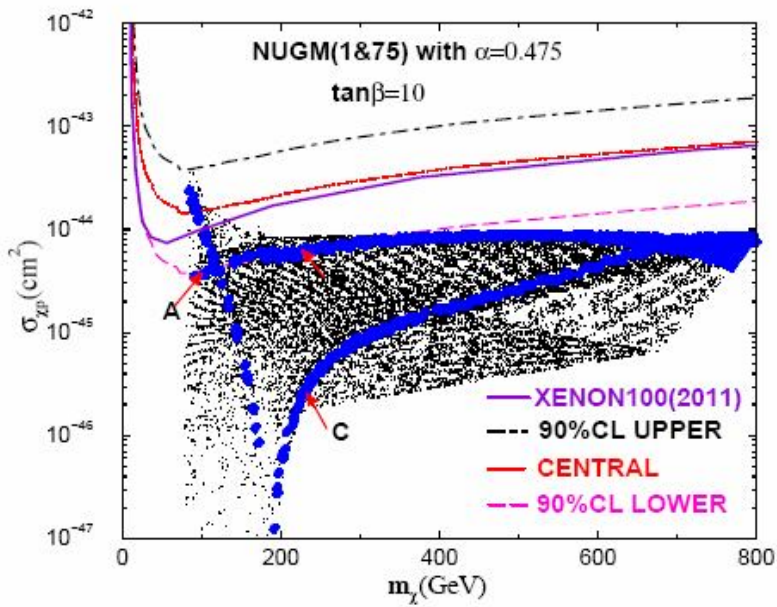

Fig 3. The direct detection cross-section prediction of the 1+75 model. The upper and lower band of blue dots represent the WMAP relic density satisfying points corresponding to the respective red bands of figure 2 [23].

The model predicts a large direct DM detection cross-section, since the dominant contribution from Higgs exchange is proportional to the product of the higgsino and gaugino components of DM. Fig 3 compares the model prediction with the experimental band corresponding to the central and 90% CL limits of the two CDMS candidate events [11]. The 90% CL upper limit of the XENON100(2011) data [12] is also shown for comparison. The upper band of WMAP relic density satisfying points are already being probed by these experiments. In fact the very recent XENON100(2012) data seems to give an upper limit very close to this upper band [13]. The SUSY mass spectra corresponding to the three representative points A,B,C of Fig 3 are listed in table 2. One clearly sees an inverted hierarchy in the squark masses, where the stop is significantly lighter than the other squarks. Thus one expects the gluino decay to proceed mainly via the stop, resulting in four b jets from the decay of the gluino pair. Therefore one can use ≥ 3 b-tags to search for the SUSY signal of this model with the available 7 TeV LHC data up to a DM mass of ~250 GeV, corresponding to a gluino mass of ~ 800 GeV [23].

| Model | $\tilde{g}$ | $\tilde{q}_L$ | $\tilde{q}_R$ | $\tilde{t}_1$ | $\tilde{b}_1$ | $\tilde{e}_l$ | $\tilde{\tau}_1$ | $\chi_1^0$ | $\chi_2^0$ | $\chi_1^+$ | $\chi_2^+$ |
|---|---|---|---|---|---|---|---|---|---|---|---|
| A | 433 | 1280 | 1274 | 759 | 1054 | 1263 | 1246 | 104 | 122 | 123 | 271 |
| B | 793 | 1480 | 1440 | 902 | 1246 | 1375 | 1327 | 227 | 256 | 257 | 501 |
| C | 722 | 750 | 660 | 483 | 649 | 437 | 237 | 231 | 301 | 302 | 490 |

Table 2. The SUSY mass spectra corresponding to the three representative points of fig 3 [23].

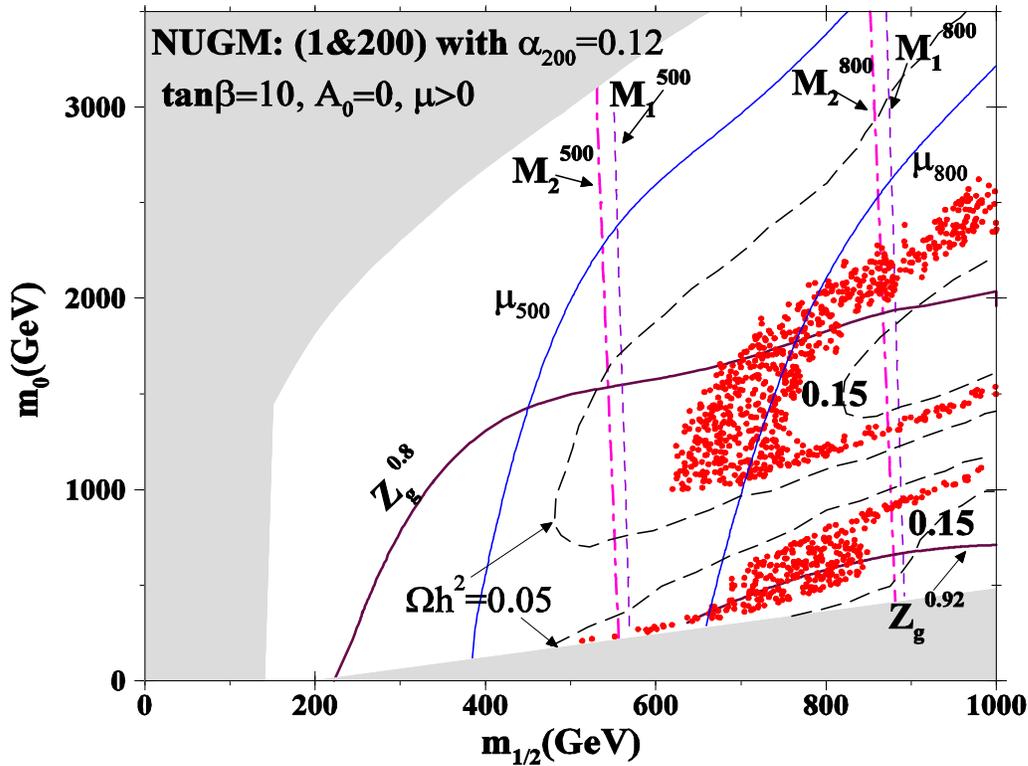

Fig 4. The composition and relic density of DM in the 1+200 model [7].

Figure 4 shows the $m_0$-$m_{1/2}$ parameter space of the 1+200 model for an optimal choice of $\alpha_{200} = 0.12$ with contours of $M_1$, $M_2$ and $\mu$. The near equality of these three masses shows that DM is a mixed neutralino state of comparable bino, wino and higgsino components. It also shows the DM relic density contours, with the WMAP relic density satisfying regions indicated by the bands of red dots. Like figure 2, the uppermost band corresponds to the annihilation of χχ into WW, ZZ and fermion-pairs (via Z), while the two lower bands together represent their resonant annihilation into heavy fermion pairs via A.

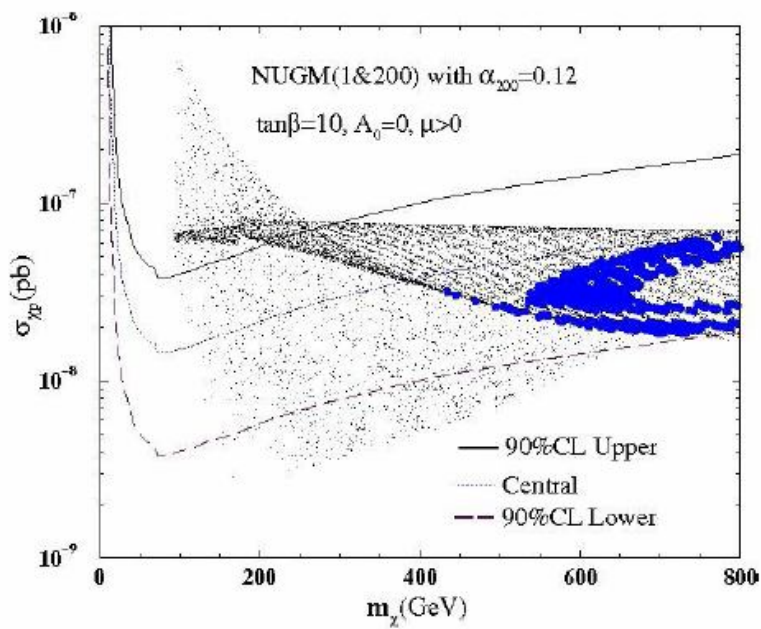

Fig 5. The predicted direct detection cross-section of the 1+200 model [24]. The blue dots represent the WMAP relic density satisfying points corresponding to the red dots of figure 4.

Fig 5 compares the 1+200 model prediction with the experimental band representing the central value and the 90% CL upper and lower limits corresponding to the two candidate events of the CDMS data [11]. The model predicts very large direct detection cross-section at the level suggested by these two candidate events. Unfortunately, the recent XENON100 (2012) data [13] gives a stringent 90%CL upper limit, roughly coinciding with the shown lower limit from the two candidate events, which is in conflict with this model prediction.

It may be added here that both these models of mixed neutralino DM predict large indirect detection signal in terms of high energy neutrinos coming from their pair annihilation inside the sun, which can be detected by the IceCube experiment [7].

4. **Wino DM in the AMSB Model**

Finally let us consider the most popular model of nonuniversal gaugino masses, i.e. the anomaly mediated SUSY breaking model [25]. Here the GUT scale gaugino masses arise from SUSY breaking in the hidden sector via the super-Weyl anomaly contributions. Being loop effects these contributions would be small compared to the tree-level contributions from eqs. (1-4). The AMSB model tacitly assumes that the SUSY breaking superfield in the hidden sector does not belong to one of the representations appearing in the symmetric product of the two adjoint representation of the GUT group in (2); so that the tree-level contributions to gagino masses are forbidden by symmetry considerations.

The GUT scale gaugino masses in the AMSB model are given in terms of the gravitino mass $m_{3/2}$ via the Callan-Symanzik β functions of their respective gauge couplings, i.e.

$$M_1^G = \frac{33}{5}\frac{\alpha_1}{4\pi}m_{3/2}, M_2^G = \frac{\alpha_2}{4\pi}m_{3/2}, M_3^G = -3\frac{\alpha_3}{4\pi}m_{3/2}. \tag{13}$$

With the universality of the gauge couplings at the GUT scale one clearly sees the wino to be the lightest gaugino. Evolving the gaugino masses down via the RGE one sees that the wino remains the lightest gaugino at the EW scale as well. Thus one has a wino DM over most of the parameter space of the AMSB model [26]. Like the higgsino, the wino can efficiently pair-annihilate via its isospin gauge coupling to W boson, leading to an under-abundance of DM relic density for sub-TeV wino mass. In fact wino has twice as large an isospin as higgsino, so that one gets the right DM relic density for a wino mass twice as large as the higgsino case, i.e. ~ 2 TeV. This is too large to be probed by LHC experiments. Besides the direct DM detection rate is severely suppressed by the large mass as well as the absence of a significant higgsino component in the DM.

By far the best chance of detecting a TeV scale wino DM lies in the indirect detection experiments looking for the products of their pair annihilation,

$$\chi\chi \xrightarrow{\tilde{\chi}_1^+} W^+W^-. \tag{14}$$

In particular the leptonic decay of W bosons can give a hard positron spectrum of the type observed by PAMELA experiment. Moreover, a possible source of the large boost factor required to match the size of the PAMELA data is provided by the Sommerfeld effect [27,28]. To understand this effect one notes that the annihilation process (14), mediated by a TeV scale charged wino exchange, has a negligibly small range of interaction. So the annihilation cross-section is proportional to the modulus square of the $\chi\chi$ wave function at negligibly small relative distance, $|\psi(0)|^2$. On the other hand, the incoming $\chi\chi$ pair has attractive interaction via a relatively long range W boson exchange which distorts the wave function, resulting in the enhancement of $|\psi(0)|^2$. One can solve the nonrelativistic scattering problem nonperturbatively by solving the Schrodinger equation. One finds that the effect is modest except for the resonance points, where one can have enhancement factors of several orders of magnitude, as shown in figure 6 [29].

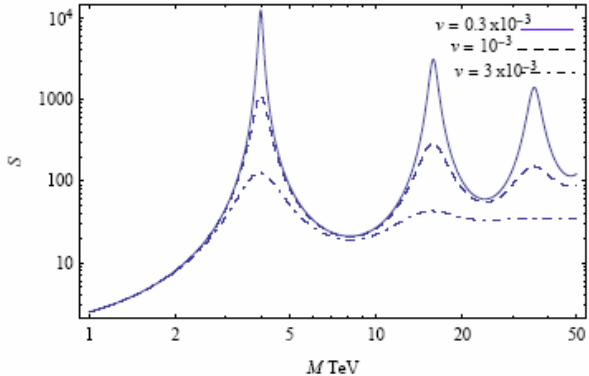

Fig 6. Sommerfeld enhancement factor from W boson exchange as a function of the DM mass at different relative velocities [29].

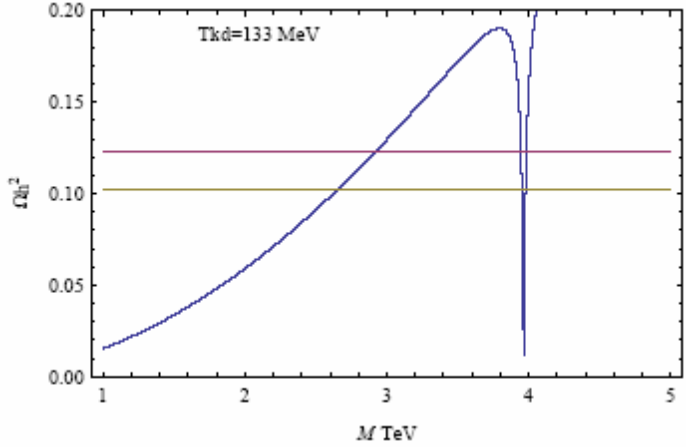

Fig 7. The suppression of the DM relic density at the Sommerfeld resonance. The horizontal band represents the WMAP relic density (12).

The Sommerfeld enhancement of the DM pair annihilation cross-section results in a significant suppression of the DM relic density at the resonance point [30], as shown in figure 7 for our wino DM [29]. Thus one can simultaneously have a large enhancement factor of the annihilation cross-section, $S \sim 10^4$, and also reconcile the DM relic density with the WMAP value (12) by choosing the wino DM mass to be very close to the first resonance value of ~ 4 TeV. The resulting spectrum of the positron flux ratio is compared with PAMELA experiment in figure 8. It may be added here that the model prediction for the anti-proton spectrum is also in reasonable agreement with the PAMELA data [29].

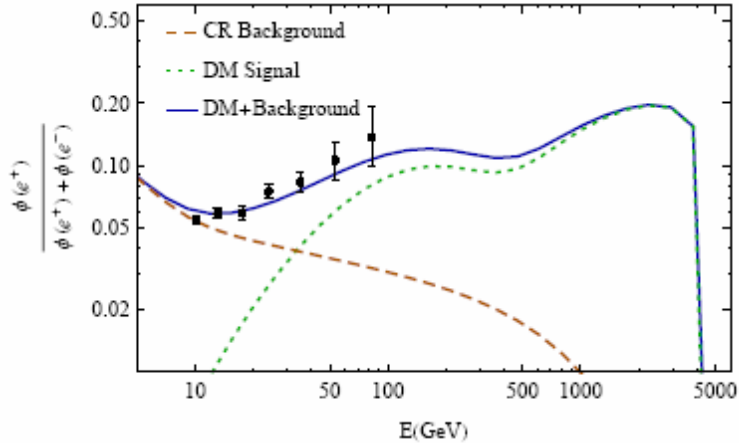

Fig 8. The predicted positron flux ratio of a 3.98 TeV wino DM is compared with the PAMELA data using a Sommerfeld enhancement factor of ~ $10^4$ [29].

## Acknowledgement:

I thank the organizers of the PASCOS 2012 and in particular Myriam Mondragon and Liliana Velasco-Sevilla for their wonderful hospitality. The works reported here were supported in part by a senior scientist fellowship of the Indian National Science Academy.